\documentclass[aps,prl,twocolumn,showpacs,superscriptaddress,groupedaddress]{revtex4-1}
\usepackage{graphicx}
\usepackage{amssymb}
\usepackage{amsmath}
\newlength{\figwidth}
\usepackage{bm}
\usepackage{xcolor}
\usepackage{color}
\usepackage{hyperref}
\hypersetup{
 colorlinks=true,
 linkcolor=blue,
 filecolor=magenta,  
 urlcolor=cyan,
}
\definecolor{airforceblue}{rgb}{0.36, 0.54, 0.66}
\usepackage{epsf}
\begin{document}
\preprint{\today}
\title{Cooperative Charge Pumping and Enhanced Skyrmion Mobility}
\author{Adel Abbout }
\affiliation{King Abdullah University of Science and Technology (KAUST),
Physical Science and Engineering Division, Thuwal 23955-6900, Saudi Arabia}

\author{Joseph Weston}
\affiliation{University Grenoble Alpes, INAC-PHELIQS, F-38000 Grenoble, France
CEA, INAC-PHELIQS, F-38000 Grenoble, France}

\author{Xavier Waintal}
\affiliation{University Grenoble Alpes, INAC-PHELIQS, F-38000 Grenoble, France
CEA, INAC-PHELIQS, F-38000 Grenoble, France}

\author{Aur\'elien Manchon}
\affiliation{King Abdullah University of Science and Technology (KAUST),
Physical Science and Engineering Division, Thuwal 23955-6900, Saudi Arabia}

\begin{abstract}

The electronic pumping arising from the steady motion of ferromagnetic skyrmions is investigated by solving the time evolution of the Schr\"odinger equation implemented on a tight-binding model with the statistical physics of the many-body problem. It is shown that the ability of steadily moving skyrmions to pump large charge currents arises from their non-trivial magnetic topology, i.e. the coexistence between spin-motive force and topological Hall effect. Based on an adiabatic scattering theory, we compute the pumped current and demonstrate that it scales with the reflection coefficient of the conduction electrons against the skyrmion. Finally, we propose that such a phenomenon can be exploited in the context of racetrack devices, where the electronic pumping enhances the collective motion of the train of skyrmions.
\end{abstract}
\pacs{
}
\maketitle


\paragraph{Introduction} - The electrical manipulation of magnetic textures has stimulated intense investigations lately, unraveling a wealth of dynamical phenomena that can be exploited towards memory and logic applications \cite{Allwood2005,Parkin2008}. Among the vast zoology of existing magnetic textures, skyrmions are regarded as promising candidates for high-density storage and logic applications \cite{Fert2013}. Magnetic skyrmions are topologically nontrivial magnetic textures characterized by a quantized topological charge \cite{Skyrme,Fert}. While skymions were originally obtained in the form of stable lattices in bulk non-centrosymmetric magnets \cite{Muhlbauer2009,Yu2010} or at metallic interfaces \cite{Heinze} at low temperature, a recent breakthrough has led to the observation of individual metastable magnetic skyrmions at room temperature at transition metal interfaces \cite{Jiang2015,Woo2016,Boulle,Moreau}. This achievement enables the electrical manipulation of individual bits of information and the in-depth investigation of current-driven skyrmion motion in thin nanowires. The superiority attributed to magnetic skyrmions compared to magnetic domain walls is related to their topology and to the energy barrier that needs to be overcome to annihilate a skyrmion \cite{Nagaosa}. This property makes them robust against certain classes of defects (point defects, edge roughness) and weakly deformable when subjected to reasonable (field or current) drive \cite{Iwasaki2013b}. As a result, skyrmions are expected to display high mobility and low critical driving current in weakly disordered media \cite{Jonietz}. Although this topological protection does not apply in strongly disordered systems, recent experiments in polycrystalline thin films have shown that skyrmions can reach mobilities as large as 100 m/s for current densities of 5$\times$10$^7$ A/cm$^2$ \cite{Woo2016}, comparable to domain walls. \par
 
Another important characteristic of magnetic skyrmions is the presence of a spin-Berry phase resulting in an emergent electromagnetic field \cite{Taguchi2001}. Indeed, the chiral configuration of the magnetic moments forming the skyrmion gets imprinted in the wavefunction of the conduction electrons such that they experience a spin-dependent effective electromagnetic field \cite{Barnes2007,Tatara}
\begin{eqnarray}\label{eq:Eem}
{\bf E}_{\rm em}^s&=&(s\hbar/2e)[{\bf m}\cdot(\partial_t{\bf m}\times\partial_i{\bf m})]{\bf e}_i,\\\label{eq:Bem}
{\bf B}_{\rm em}^s&=&-\epsilon_{ijk}(s\hbar/2e)[{\bf m}\cdot(\partial_i{\bf m}\times\partial_j{\bf m})]{\bf e}_k,
\end{eqnarray}
where ${\bf m}$ is the unit magnetization vector, $\epsilon_{ijk}$ is Levi-Cevita symbol, ${\bf e}_i$ is the $i$-th unit vector in cartesian coordinates and $s=\pm1$ accounts for spin projection on ${\bf m}$. When the skyrmion is static, ${\bf E}_{\rm em}^s\rightarrow0$ and the emergent magnetic field ${\bf B}_{\rm em}^s$ is responsible for the deviation of the flowing electron trajectory, resulting in topological charge and spin Hall effects \cite{Nagaosa,Ndiaye2017}, as well as topological torque \cite{Bisig2016,Akosa2017}. As a reaction, the skyrmion experiences a Magnus force that pushes it sideway, as observed experimentally \cite{Jiang2017,Litzius2017,Legrand2017}. On the other hand, when the skyrmion is moving, ${\bf E}_{\rm em}^s\neq0$, and it is expected to pump spin and charge currents through the so-called spin-motive force, an effect experimentally reported in quasi-one dimensional magnetic domain walls (1D-DW) \cite{Yang2009,Hayashi2012}. In 1D-DW though, the charge pumping only occurs in the {\em turbulent} regime of motion, such that $\langle{\bf E}_{\rm em}^s\rangle=\langle{\bf m}\cdot(\partial_t{\bf m}\times\partial_i{\bf m})\rangle\neq0$, and vanishes during {\em steady} motion (also known as the flow regime).

 
In this Letter, we uncover the interplay between spin currents and skyrmion dynamics by directly solving the time-dependent Schr{\"o}dinger equation using state-of-the-art numerical tools \cite{Joseph,Gaury}. This approach enables us to describe the various pumping phenomena beyond the adiabatic limit \cite{Iwasaki2013b}. We show that, in sharp contrast to 1D-DW, magnetic skyrmions moving steadily along narrow nanowires pump a sizable charge current that can be detected experimentally. We propose to take advantage of these phenomena in a magnetic racetrack and demonstrate that mutual charge pumping in a train of skyrmions results in enhanced collective mobility.

 \paragraph{Model and method} - We consider a skyrmion in a perpendicularly magnetized waveguide driven by an external force, and having a velocity $v$ confined along the x-axis. In other words, Magnus force is cancelled due to confinement. Since our interest lies on the electronic pumping induced by the skyrmion motion, the physical origin of the external force that drives the motion is unimportant and can be a non-adiabatic torque \cite{Fert2013}, a topological torque \cite{Akosa2017} or spin Hall effect arising from an adjacent heavy metal \cite{Woo2016}. Finally, we also assume that the skyrmion remains stable and rigid throughout its motion. Slight deformations of the skyrmion in the presence of weak disorder is not expected to modify the present results qualitatively. \par
 
 Within these assumptions, the real-space tight-binding Hamiltonian reads
 \begin{equation}\label{eq:H}
 \hat{\mathcal{H}}(t)= -\gamma \sum_{\langle i,j \rangle} \hat{c}_i^\dagger \hat{c}_j- \Delta \sum_i {\bf{m}}_i(t)\cdot \big( \hat{c}_i^\dagger \hat{\bf{\sigma}} \hat{c}_i \big)+ \hat{V}_\text{conf},
 \end{equation}
 where $\hat{c}_i=(c_i^\uparrow,c_i^\downarrow)$ is the annihilation operator for electrons with spin up and spin down on site $i=(x,y)$ and $\hat{c}_i^\dagger $ is the corresponding creation operator. $\Delta$ is the exchange parameter, $\gamma$ is the hopping coefficient and the sum $\langle i,j \rangle$ is only taken over nearest neighbors. The magnetic texture is described by the set of the magnetic moments ${\bf{m}}_i(t)$ which takes the form of a skyrmion with a center linearly translating in time in the x-direction at a velocity $v$. Thus, the dynamical texture reads
 \begin{equation}
{\bf m}=\left(\sqrt{1-l^2(t)} \sin\phi(t),\sqrt{1-l^2(t)} \cos\phi(t),l(t)\right),
 \end{equation}
 with $l(t)=-\cos(\pi r(t)/R)$ and $\sin\phi(t)= \big(x-(x_0+v t)\big)/r(t)$ and $r(t)=\sqrt{\big(x-(x_0+v t)\big)^2+(y-y_0)^2}$. $R$ is the radius of the skyrmion and $(x_0,y_0)$ is the initial position of the skyrmion. $\hat V_\text{conf}$ is the waveguide confining potential (hard walls). Equation \eqref{eq:H} is implemented on a tight-binding model consisting of a magnetic nanowire connected to two (left and right) semi-infinite leads. In the present work, no bias voltage is applied across the system so that the currents computed only stem from skyrmion-induced electronic pumping.\par

The main task in computing the different physical observables is to solve the time-dependent Schr{\"o}dinger equation and to determine the different time-dependent propagating wave functions $\Psi_{\alpha E}(t)$ coming from each lead $\alpha$ at energy $E$. The total current flowing through a cross-section transverse to the waveguide direction reads 
 \begin{equation}\label{eq:current}
 I(t)=\sum_{\alpha,ij} \int \frac{dE}{2\pi} f_\alpha(E) I_{\alpha,ij}(E,t),
 \end{equation}
 where $f(E)$ is Fermi-Dirac distribution, and $I_{\alpha,ij}(E,t)$ is the current contribution of electrons coming from lead $\alpha$ at energy $E$ between sites $i$ and $j$ \cite{Joseph,Gaury}, 
 $I_{\alpha,ij}(E,t)=2\Im\big\{ [\Psi_{\alpha E}^\dagger(t)]_i \mathcal{H}_{ij}(t) [\Psi_{\alpha E}(t)]_j \big\}$. Here, $\Im\big\{...\big\}$ denotes the imaginary part. Equation \eqref{eq:current} shows that the physics of dynamical Hamiltonians is in general governed by the whole Fermi sea and not just by the Fermi surface although, as discussed below, the restriction to low skyrmion velocities makes the contribution of the Fermi surface dominating. The summation in Eq. \eqref{eq:current} includes the $M$ propagating modes in the waveguide of width $W$ ($M=\frac{2W}{\lambda}$, $\lambda$ being the electron wavelength).  \par

 \begin{figure}[t]
\includegraphics[width=6cm]{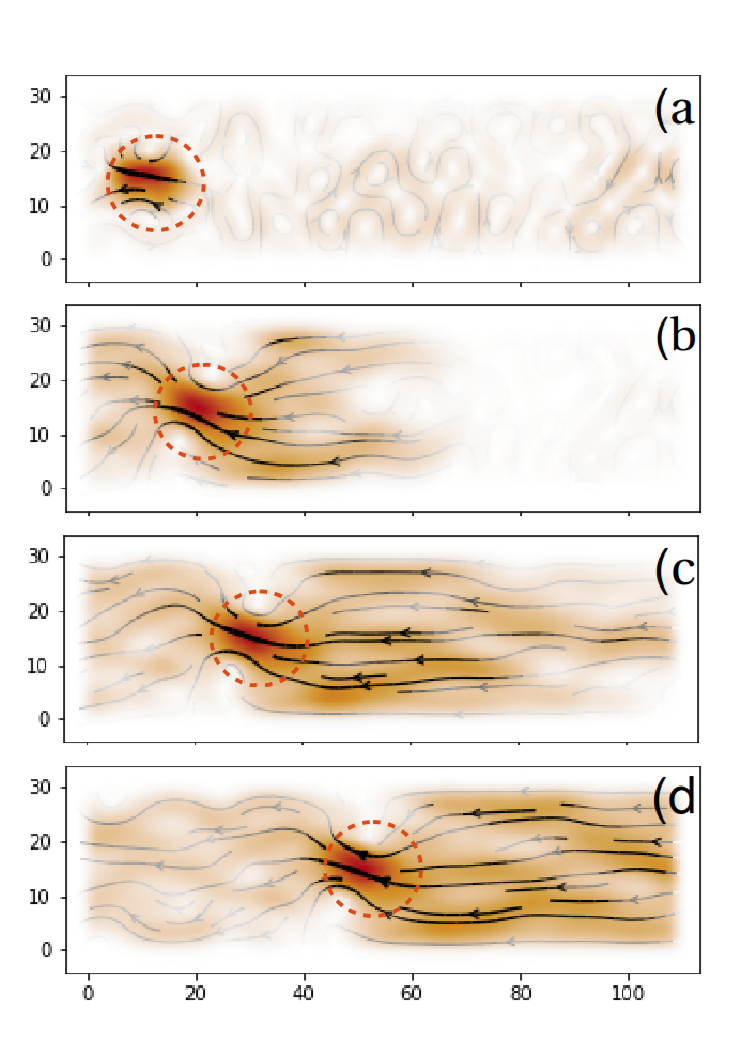}
\centering
\caption{(Color online) 2D map of the current lines at different times: (a) $t=4t_0$, (b) $50 t_0$, (c) $100 t_0$, and (d) $200 t_0$. The black arrows denote the direction of the charge current, and the color shading amounts for the current gradient. The dark shaded region reveals the position of the skyrmion, where the current gradient is maximum. The parameters are $v=0.2a/t_0$, $W=31a$, $2R=15a$ and $E_F=-3.8\gamma$. Initial skyrmion position is $x(0)=15a$.}
\label{Current_lines}
\end{figure}

\paragraph{Skyrmion-induced electronic pumping} - We now consider a quasi-1D waveguide with a width $W=31a$ in which a skyrmion of diameter $2R=15a$ is moving with a velocity $v=0.2 a/t_0$, where $a$, $t_0$ are the lattice and time subdivisions respectively \cite{Joseph3}. The exchange parameter is set to $\Delta=0.1\gamma$. The skyrmion is set into motion by ramping its velocity from zero to a steady value $v$ over a ramping time of $\sim$3$t_0$. This ramping time is kept short to minimize its impact on electronic pumping. A two-dimensional (2D) map of the charge current induced by the skyrmion motion is plotted in Fig. \ref{Current_lines} for various times. In the transient regime close to $t=0$, current vortices reveal the inhomogeneity of the wavefunction due to the presence of the skyrmion [Fig. \ref{Current_lines}(a)]. At longer times, a stationary regime establishes progressively Fig. \ref{Current_lines}(b)-(d). The streamlines show that the pumping is asymmetric despite the axial symmetry of the system, unveiling the presence of topological Hall effect.

 \begin{figure}[th]
\includegraphics[width=9cm]{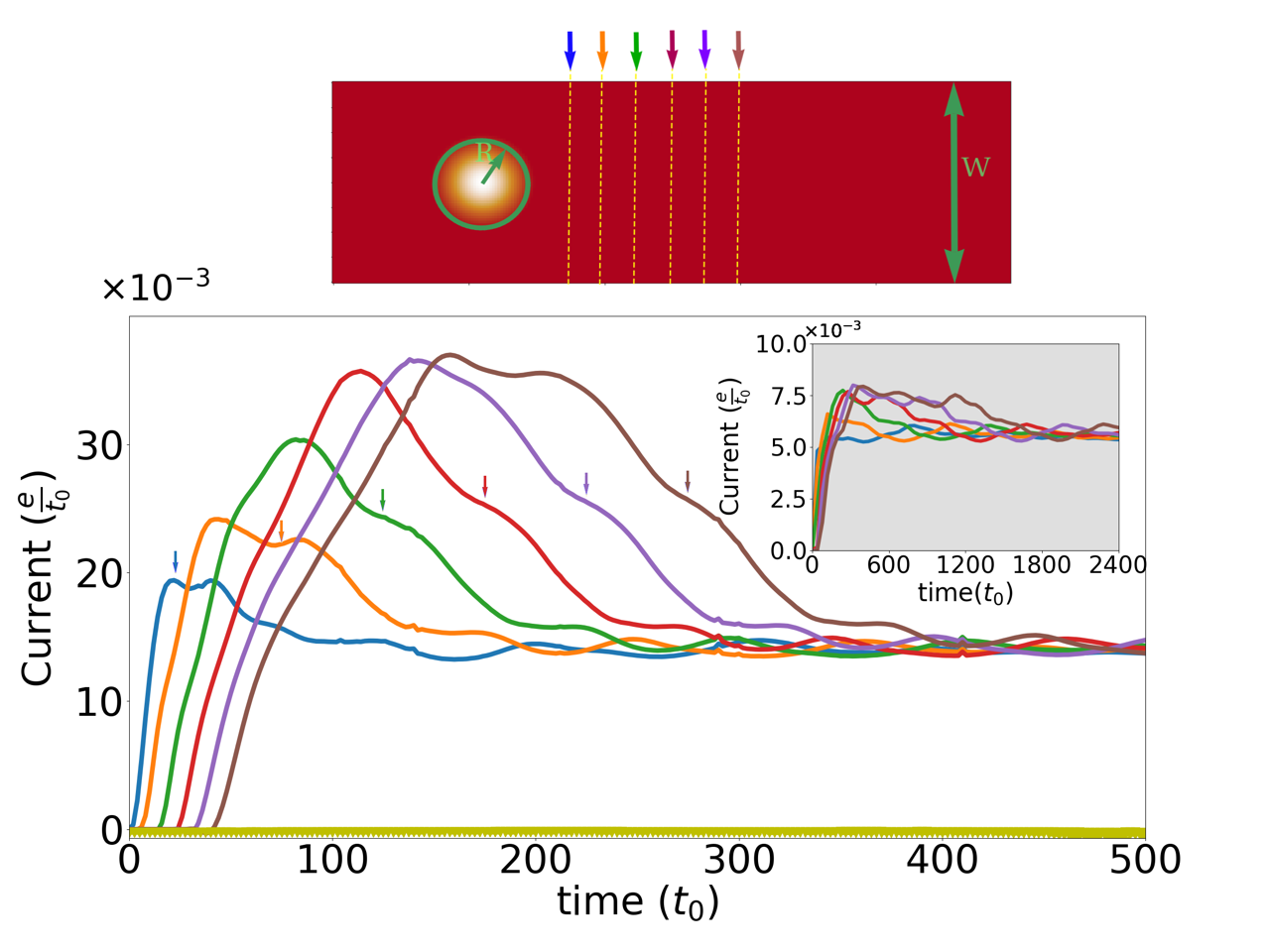}
\centering
\caption{(Color online) Trace of the charge current computed at different interfaces in the waveguide. Each curve corresponds to the current probed at a specific interface, as indicated on the top sketch (dashed vertical lines). The colored arrows indicate the time when the skyrmion center crosses a specific interface. These interfaces are spaced by a distance of 10$a$ and the parameters are $W=51a$, $2R=15a$, $v=0.2 a/t_0$ and $\Delta=0.1\gamma$. The symbols $ \begingroup
 \color{yellow}
 \blacktriangledown
 \endgroup $ denote the current pumped by a 1D-DW with the same parameters. The inset is obtained for $v=0.05 a/t_0$.}
\label{Pumped_Current}
\end{figure}

The total current pumped by the moving skyrmion is reported on Fig. \ref{Pumped_Current} as a function of time and probed at different positions along the waveguide, as depicted in the top panel. After a sharp increase at early times due to the fast redistribution of the electronic density upon the (slow) skyrmion motion, the pumped current reaches a maximum and then decreases towards a steady value. The visible small ripples in the stationary current come from the fact that the skyrmion's shape changes slightly during motion because of the lattice discretization. The different magnitude recorded at different cross-section reflects the progressive onset of the steady regime for electronic pumping. The inset of Fig. \ref{Pumped_Current} shows the same calculation for a much slower skyrmion velocity, $v=0.05 a/t_0$. Remarkably, the onset of steady electronic pumping takes much longer to stabilize ($\sim$1800$t_0$ compared to $\sim$400$t_0$ in the main panel), while the magnitude of the pumped current is smaller ($\sim 0.55\times10^{-2} e/t_0$ compared to $\sim 1.4\times10^{-2} e/t_0$). \par

For the sake of completeness, we also computed the current pumped by a 1D-DW, depicted by the flat line in Fig. \ref{Pumped_Current}. In 1D-DW, the current pumped by a moving texture is directly proportional to the emergent electric field, ${\bf E}_{\rm em}^s$. In the steady motion regime, $\partial_t{\bf m}=-v\partial_x{\bf m}$, such that ${\bf E}_{\rm em}^s=-v(s\hbar/2e)[{\bf m}\cdot(\partial_x{\bf m}\times\partial_y{\bf m})]{\bf y}$. Hence, because ${\bf E}_{\rm em}^s\|{\bf y}$, there is no spin-motive force along the axis of the waveguide, and the pumped current vanishes, consistently with the experimental observations \cite{Yang2009,Hayashi2012}. In contrast, a skyrmion is a 2D object such that the pumped current arises from the cooperation between the spin-motive force ($\sim{\bf E}_{\rm em}^s$) and the topological Hall effect ($\sim{\bf B}_{\rm em}^s$). Indeed, in the adiabatic limit the semiclassical spin-dependent pumped current reads $I^s=-\int d{\cal S}\sigma^s_{\rm H}{\bf E}_{\rm em}^s\times{\bf B}_{\rm em}^s$ \cite{Akosa2017}, where ${\cal S}$ is the waveguide area and $\sigma^s_{\rm H}$ is the ordinary Hall conductivity for spin $s$. As a result, the total charge current is
\begin{equation}\label{eq:Ipump}
I=-v(\hbar/2e)^2\int d{\cal S}(\sigma^\uparrow_{\rm H}+\sigma^\downarrow_{\rm H})[{\bf m}\cdot(\partial_x{\bf m}\times\partial_y{\bf m})]^2.
\end{equation}
Hence, the pumped current is proportional to the skyrmion velocity and does not vanish because of the cooperation between spin-motive force and topological Hall effect, a unique feature of topologically non-trivial magnetic textures.\par

Expression \eqref{eq:Ipump} only applies in the adiabatic and semiclassical limit (large number of modes). To provide a more accurate estimate of the current pumped by the skyrmion, we follow a scattering matrix approach \cite{Buttiker}. On the basis of the Redheffer product for combining scattering matrices \cite{Redheffer}, we express the "frozen" scattering $S$ matrix as a function of the position of the skyrmion. After some algebra \cite{SupplementalMaterial}, we find $S= QS^0 Q$ with the elements of the diagonal matrix $Q$ being $Q_{\alpha\alpha}=\exp{(i k_\alpha v t )}$ if $ \alpha\le M$ (rightward modes) and $Q_{\alpha\alpha}=\exp{(-i k_\alpha v t)}$ if $ \alpha> M$ (leftward modes). Here, $k_\alpha$ is the momentum vector of mode $\alpha$ and $S^0$ is the initial scattering matrix. With this, we get $S_{\alpha\beta} = S_{\alpha\beta}^0 \exp{(\pm i k_{\alpha\beta} v t)} $ with $k_{\alpha\beta}=k_\alpha\pm k_\beta$. The $\pm$ sign depends on whether the modes $\alpha$ and $\beta$ are from the same lead or not \cite{SupplementalMaterial}. At long times, when the stationary regime establishes, we find the out of equilibrium Fermi distribution $f^\text{out}_\alpha=\sum_\beta |S^0_{\alpha\beta}|^2 f_{0}(E+\hbar k_{\alpha\beta}v) $, $f_0$ being the distribution at equilibrium. At this stage, the stationary current can be expressed in an energy integral form \cite{Buttiker} and then estimated numerically. In order to obtain an analytical expression, we restrict ourselves to the case of slow skyrmion although our algorithm goes beyond this restriction. Indeed, the skyrmion velocity ($\sim$ 10-100 m/s) remains much smaller than the electronic velocity ($\sim$ 10$^5$ m/s). Therefore, if the electrons flow through the skyrmion before the texture moves substantially, one can consider the scattering as adiabatic and thus the use of the frozen $S$ matrix is justified \cite{Levinson}. This condition is fulfilled when $\hbar k_{\alpha\beta} v\ll E_F $ $\forall \{\alpha,\beta\}$, or equivalently $2 \hbar k_F v\ll E_F$, where $k_F$ and $E_F$ are the Fermi wavenumber and energy, respectively.

\begin{figure}[t]
\includegraphics[width=7cm]{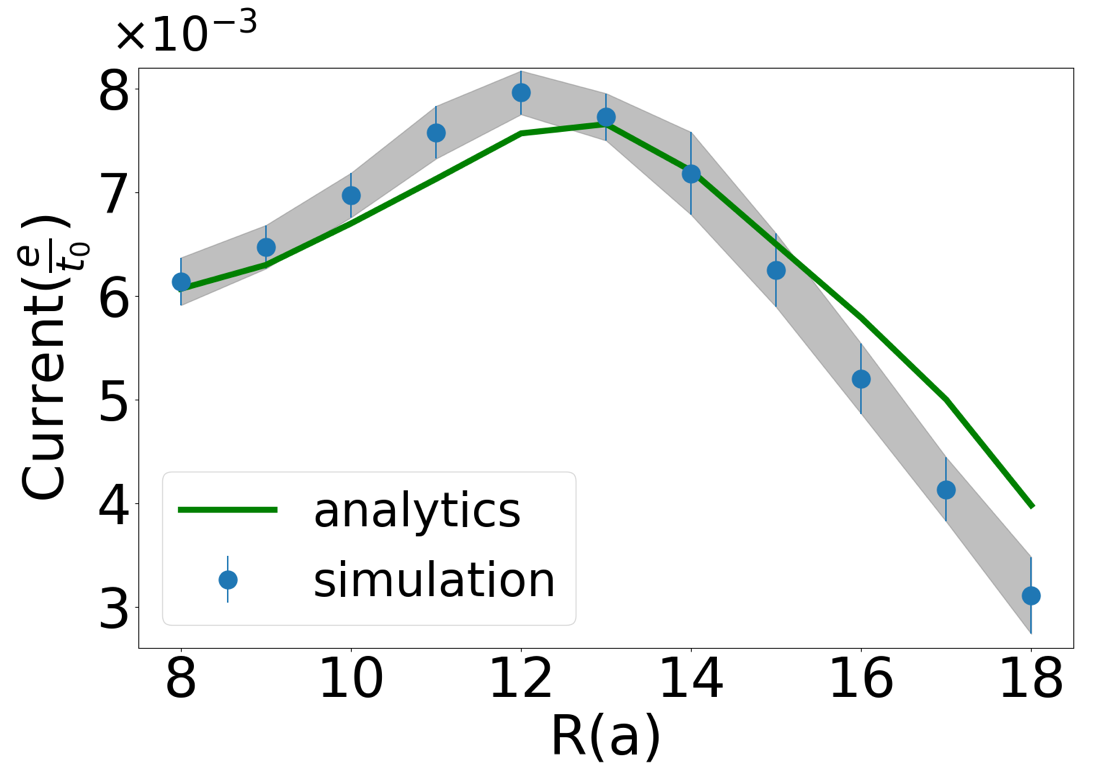}
\centering
\caption{(Color online) Stationary current pumped by one skyrmion as a function of its radius as computed numerically (blue symbols) and analytically (green solid line). The grey region is the range of the fluctuation due to lattice effects (i.e., the ripples in Fig. \ref{Pumped_Current}). The velocity is $v=0.07a/t_0$ and the other parameters are the same as in Fig. \ref{Current_lines}.}
\label{Current}
\end{figure}

Within this adiabatic behavior, only electrons around the Fermi energy are pertinent in the expression of the stationary current and all the contributions from the Fermi sea cancel. At this stage, the expression of the current boils down to the calculation of the emissivity into the contact $m$ defined as $\frac{-e}{2 \pi v} \sum _\beta \sum_{\alpha \in m} \Im \{S_{\alpha\beta}^\ast\partial_t S_{\alpha \beta}\}$ \cite{Buttiker2, Brouwer}. Within this approximation, the charge current reads,
\begin{equation}\label{currentanalytic}
 I=-\frac{e}{\pi} v \overline{k} \mathcal{R},
\end{equation}
 $\overline{k}$ is the barycenter of $k_\alpha$ with coefficients $\sum_{\beta \in m}|r_{\alpha\beta}|^2$ and $\mathcal{R}$ is the total reflection of the system. This expression reflects the rowing effect of the skyrmion. Indeed, the pumped current can be enhanced by increasing the velocity of the skyrmion (see Fig. \ref{Pumped_Current}), taking a larger Fermi energy (faster electrons) or considering a more reflective skyrmion. For instance, let us consider a skyrmion moving with a velocity $v=0.05 a/t_0$ in a waveguide of width $W=51a$ (inset of Fig. \ref{Pumped_Current}). Assuming $a=1~\text{nm}$, $t_0=2.5~\text{ ps}$, we obtained $v=20 $ m/s, $\overline{k}=0.2~\text{nm}^{-1}$ and $\mathcal{R}=1.8$, and Eq. \eqref{currentanalytic} gives a charge current $I=0.366$ nA, which is in good agreement with the value computed numerically in the inset of Fig. \ref{Pumped_Current}, $I=0.352$ nA. If we consider a thickness of $d=0.5 \text{ nm}$ for a magnetic thin film, we get a current density $j_p=1.43 \times 10^7$ A/m$^2$, which is accessible experimentally. This value can be considerably increased by increasing the number of conducting modes and the reflection coefficient. Figure \ref{Current} displays the pumped current computed numerically (blue symbols) and analytically (green solid line) as a function of the skyrmion radius, showing a very good agreement between the two methods. The electronic pumping reaches a maximum around $R\sim12~a$, which corresponds to the smallest wavelength of the flowing electrons, estimated around $\lambda\approx11~a$ \cite{SupplementalMaterial}. When the radius becomes larger than this wavelength, the reflection of the electrons against the skyrmion decreases and the pumping efficiency is subsequently reduced. \par

\begin{figure}[t]
\includegraphics[width=7cm]{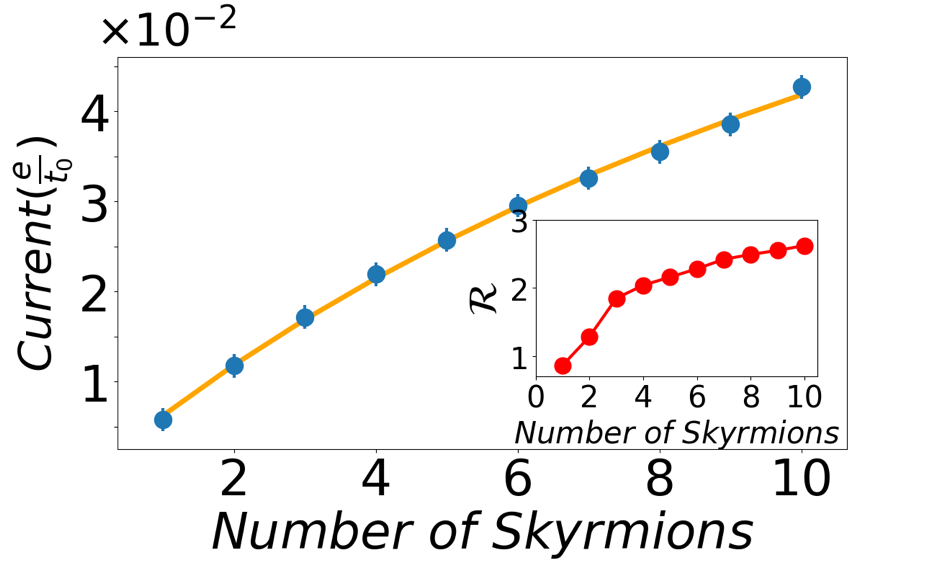}
\centering
\caption{(Color online) Pumped current as a function of the number of skyrmions. The symbols {\large {\color{airforceblue}$\bullet$}} represent the current computed numerically and the orange line is the fit with the equation for $N$ skyrmions given in the main text. The inset display the reflection coefficient as a function of the number of skyrmions. The parameters are $W=31a$. $\Delta=0.1\gamma$, $v=0.15a/t_0$ and $R=7.5a$, $E_F=-3.5\gamma$.}
\label{Current_vs_number_Skyrmion}
\end{figure}

\paragraph{Cooperative pumping and enhanced motion} - The electronic pumping can be enhanced by increasing the number of skyrmions in the system and thereby the reflection of flowing electrons. This effect is illustrated in Fig. \ref{Current_vs_number_Skyrmion}, where the pumped charge current is reported as a function of the number of skyrmions present in the nanowire. The inset shows the analog behavior for the reflection coefficient. This cooperative pumping is expected to saturate at large number of skyrmions due to the bound of $\mathcal{R}$. For instance, if we consider $E_F=-0.97\gamma$, $\Delta=0.98\gamma$ and $10$ skyrmions, we obtain a current density $50$ times larger. A crude estimate of the cooperative pumping can be obtained in the frame of a chain of one-dimensional scatterers. Assuming that each skyrmion behaves like a one-dimensional scatterer with an effective reflection coefficient $\overline{\mathcal{R}}$, a chain of $N$ skyrmions produces a total pumped current of $ I_N =-\frac{e}{\pi} v \overline{k} \frac{M}{1+(M-\overline{\mathcal{R}})/(N\overline{\mathcal{R}}) }$. In the limit of $N\rightarrow+\infty$, $I_\infty\rightarrow -\frac{e}{\pi} v \overline{k}M$. Figure \ref{Current_vs_number_Skyrmion} shows a very good test of this formula and describes well the total current as a function of the number of skyrmions. \par

In a skyrmion racetrack \cite{Fert2013}, the current pumped collectively by the train of skyrmions exerts a spin transfer torque on each individual skyrmion, enhancing the effective force exerted on each skyrmion. Consider a chain of skyrmions driven by either spin transfer or spin-orbit torque. Its steady state velocity reads $v=\chi j_0+\eta j_p$, where $j_{0(p)}$ is the injected (pumped) current density, $\chi$ is the (either spin-orbit or spin transfer) torque efficiency, and $\eta$ is the spin transfer torque efficiency. Since the pumped spin current itself depends on the velocity, $j_p=\xi v$, one obtains the renormalized velocity of the skyrmion train, $v=\chi j_0/(1-\eta\xi)$. In other words, the electronic pumping enhances the collective skyrmion velocity. The typical skyrmion mobility is 100 m/s for a current density of $10^{12}$ A/m$^2$ \cite{Woo2016}, which gives $\eta=10^{-10}$ m$^3\cdot$s/$A$. Following the estimation given above, the pumping efficiency is about $\xi=j_p/v=-e\overline{k}M/(\pi Wd)\approx-ek_F^2/\pi^2$. Here we assumed a large number of modes such that $M/W\approx k_F/\pi$ and $\overline{k}\approx k_F$. Taking $k_F=10$ nm$^{-1}$ and a ferromagnetic thickness $d=0.5$ nm, we obtain $\xi=3\times10^9$ A/(m$^3\cdot$s). This provides a renormalization of $1-\eta\xi\approx0.68$, which indicates that the mobility of the train of skyrmion can be substantially enhanced to almost 50\%.

\paragraph{Conclusion} - We showed that under steady motion, magnetic skyrmions are able to pump a charge current due to the cooperation of the spin-motive force and topological Hall effect. We propose that the cooperative electronic pumping arising from a train of skyrmions in a racetrack enhances the collective skyrmion mobility. Notice that each moving skyrmion also pumps a non-equilibrium spin density locally that enhances the magnetic damping but does not affect our conclusions \cite{Akosa2017}. This prediction calls for experimental verification and has the potential to foster the development of skyrmion racetracks \cite{Parkin2008,Fert2013}.

\acknowledgments
A. A. and A. M. acknowledge financial support from the King Abdullah University of Science and Technology
(KAUST). We acknowledge computing time on the supercomputers SHAHEEN at KAUST Supercomputing Centre and the team assistance. A. A. thanks A. Salimath, P. B. Ndiaye, S. Ghosh and C. A. Akosa for useful discussions. 
\bibliography{LETTRE-AFMP}

\begin{thebibliography}{60}

\bibitem{Allwood2005} D.A. Allwood, G. Xiong, C.C. Faulkner, D. Atkinson, . Petit, and R.P. Cowburn, \href{http://science.sciencemag.org/content/309/5741/1688.long}{Science {\bf 3009}, 1688 (2005).}
\bibitem{Parkin2008} S.S.P. Parkin, M. Hayashi, L. Thomas, \href{http://science.sciencemag.org/content/320/5873/190}{Science {\bf 320}, 190 (2008).}
\bibitem{Fert2013} A. Fert, V. Cros and J. Sampaio. \href{https://www.nature.com/articles/nnano.2013.29}{Nat. Nanotechnol. {\bf8}, 152-156 (2013).}
\bibitem{Skyrme} T. H. R. Skyrme, A unified field theory of mesons and baryons. \href{http://www.sciencedirect.com/science/article/pii/0029558262907757}{Nucl. Phys. {\bf31}, 556-569 (1962).}
\bibitem{Fert} A. Fert, N. Reyren, and V. Cros, \href{https://www.nature.com/articles/natrevmats201731}{ Nat. Rev. Mater. {\bf2}, 17031 (2017).}
\bibitem{Muhlbauer2009} S. M\"uhlbauer, B. Binz, F. Jonietz, C. Pfleiderer, A. Rosch, A. Neubauer, R. Georgii, P. B\"oni, \href{http://science.sciencemag.org/content/323/5916/915.long}{Science {\bf 323}, 915 (2009).}
\bibitem{Yu2010} X.Z. Yu et al., \href{http://www.nature.com/articles/nature09124}{Nature {\bf 465}, 901 (2010).}
\bibitem{Heinze} S. Heinze, et al. \href{http://www.nature.com/nphys/journal/v7/n9/full/nphys2045.html}{Nat. Phys. {\bf7}, 713-718 (2011).}
\bibitem{Jiang2015} W. Jiang et al., \href{https://www.ncbi.nlm.nih.gov/pubmed/26067256}{Science {\bf 349}, 283 (2015).}
\bibitem{Woo2016} S. Woo, et al. \href{http://www.nature.com/nmat/journal/v15/n5/full/nmat4593.html}{ Nat. Mater. {\bf15}, 501-506 (2016).}
\bibitem{Boulle} O. Boulle, et al. \href{http://www.nature.com/nnano/journal/v11/n5/full/nnano.2015.315.html}{ Nat. Nanotechnol. {\bf11}, 449-454 (2016).}
\bibitem{Moreau} C. Moreau-Luchaire, et al. \href{http://www.nature.com/nnano/journal/v11/n5/full/nnano.2015.313.html}{ Nat. Nanotechnol. {\bf11}, 444-448 (2016).}
\bibitem{Nagaosa} N. Nagaosa and Y. Tokura, \href{http://www.nature.com/nnano/journal/v8/n12/full/nnano.2013.243.html}{ Nat. Nanotechnol. {\bf8}, 899 (2013).}
\bibitem{Iwasaki2013b} J. Iwasaki, M. Mochizuki, and N. Nagaosa, \href{http://www.nature.com/articles/nnano.2013.176}{Nat. Nanotechnol. {\bf 8}, 742 (2013).}
 \bibitem{Jonietz} F. Jonietz et al., \href{http://science.sciencemag.org/content/330/6011/1648.long}{Science {\bf330}, 1648 (2010).}
 
\bibitem{Taguchi2001} Y. Taguchi, Y. Oohara, H. Yoshizawa, N. Nagaosa, and Y. Tokura, \href{http://science.sciencemag.org/content/291/5513/2573.long}{Science {\bf 291}, 2573 (2001).}
\bibitem{Barnes2007} S. E. Barnes, S. Maekawa, \href{https://journals.aps.org/prl/abstract/10.1103/PhysRevLett.98.246601}{Physical Review Letters {\bf98}, 246601 (2007).}
\bibitem{Tatara} G. Tatara, H. Kohno, and J. Shibata. \href{http://www.sciencedirect.com/science/article/pii/S0370157308002597}{Phys. Rep. {\bf468}, 213 (2008).}
\bibitem{Ndiaye2017} P. B. Ndiaye, C. A. Akosa, and A. Manchon, \href{https://journals.aps.org/prb/abstract/10.1103/PhysRevB.95.064426}{Phys. Rev. B {\bf95}, 064426 (2017).}
\bibitem{Bisig2016} A. Bisig et al., \href{https://journals.aps.org/prl/abstract/10.1103/PhysRevLett.117.277203}{Phys.Rev. Lett. {\bf 117}, 277203 (2016)}
\bibitem{Akosa2017} C. A. Akosa, P. B. Ndiaye, and A. Manchon, \href{https://journals.aps.org/prb/pdf/10.1103/PhysRevB.95.054434}{Phys. Rev. B {\bf95}, 054434 (2017).}
\bibitem{Stone} M. Stone, \href{ https://journals.aps.org/prb/pdf/10.1103/PhysRevB.53.16573}{Phys. Rev. B {\bf53},16573 (1996).}
 \bibitem{Jiang2017} W. Jiang et al., \href{http://www.nature.com/articles/nphys3883}{Nat. Phys. {\bf 13}, 162 (2017).}
 \bibitem{Litzius2017} K. Litzius et al., \href{https://www.nature.com/articles/nphys4000}{Nat. Phys. {\bf 13}, 170 (2017).}
 \bibitem{Legrand2017} W. Legrand et al., \href{https://pubs.acs.org/doi/abs/10.1021/acs.nanolett.7b00649}{Nano Letters {\bf 17}, 2703 (2017).} 
 \bibitem{Yang2009} S. Yang et al., \href{https://journals.aps.org/prl/abstract/10.1103/PhysRevLett.102.067201}{Phys. Rev. Lett. {\bf102} 067201 (2009).}
 \bibitem{Hayashi2012} M. Hayashi, \href{https://journals.aps.org/prl/abstract/10.1103/PhysRevLett.108.147202}{Phys. Rev. Lett. {\bf 108}, 147202 (2012).}
\bibitem{Joseph} J. Weston and X. Waintal. \href{https://journals.aps.org/prb/abstract/10.1103/PhysRevB.93.134506}{Phys. Rev. B {\bf93}, 134506 (2016).}
\bibitem{Gaury} B. Gaury, J. Weston, M. Santin, M. Houzet, C. Groth, X. Waintal . \href{https://doi.org/10.1016/j.physrep.2013.09.001}{Phys. Rep. {\bf534}, 1 (2014).}

\bibitem{Joseph3} J. Weston and X. Waintal. \href{https://link.springer.com/article/10.1007/s10825-016-0855-9}{J. Comput. Electron. {\bf15}, 1148 (2016).}
\bibitem{Buttiker} M. Moskalets and M. B{\"u}ttiker.\href{https://journals.aps.org/prb/abstract/10.1103/PhysRevB.66.035306}{Phys. Rev. B {\bf66}, 035306.}
\bibitem{Redheffer} R. Redheffer. \href{http://onlinelibrary.wiley.com/doi/10.1002/sapm19624111/epdf?r3_referer=wol&tracking_action=preview_click&show_checkout=1&purchase_referrer=www.google.com.sa&purchase_site_license=LICENSE_EXPIRED}{J. Math. Phys., 41 (1962).}
\bibitem{SupplementalMaterial} Supplementary Materials.
\bibitem{Buttiker2} M. B{\"u}ttiker, H. Thomas, and A. Pr\^{e}tre, \href{https://link.springer.com/article/10.1007/BF01307664}{Z. Phys. B {\bf94}, 133 (1994).}
\bibitem{Brouwer} P. W. Brouwer. \href{https://journals.aps.org/prb/abstract/10.1103/PhysRevB.58.R10135}{Phys. Rev. B {\bf58}, R10135(R).} 
\bibitem{Levinson} O. E. Wohlman, A. Aharony and Y. Levinson. \href{https://journals.aps.org/prb/abstract/10.1103/PhysRevB.65.195411}{Phys. Rev. B {\bf65}, 195411 (2002).}
\end{thebibliography}


\end{document}